\documentclass[11pt]{article}
\pdfoutput=1
\usepackage[english]{babel}
\usepackage[numbers]{natbib}
\usepackage[T1]{fontenc}
\usepackage[utf8]{inputenc}
\usepackage{amsmath,braket,amsthm,amssymb}
\usepackage{graphicx}
\usepackage{tikz}
\usepackage{subcaption}
\usepackage[margin=1in]{geometry}
\usepackage{hyperref}
\usepackage{cleveref}
\usepackage{xcolor}
\usepackage{authblk}
\usepackage{comment}
\usepackage[toc,page]{appendix}
\usepackage{algorithm, setspace}
\usepackage{algpseudocode}
\usepackage{microtype}
\microtypesetup{nopatch={footnote}}
\usepackage{thmtools}
\usepackage{thm-restate}

\newcommand{\semigeq}{\succeq}

\renewcommand\footnotemark{}

\newtheorem{theorem}{Theorem}

\newtheorem{definition}[theorem]{Definition}

\newcommand{\tr}{\mathrm{Tr}}

\newcommand{\knote}[1]{\footnote{{\color{red} {\bf Kevin}: {#1}}}}
\newcommand{\onote}[1]{{\color{blue} {\bf Ojas}: {#1}}}

\renewcommand{\knote}[1]{}
\renewcommand{\onote}[1]{}

\title{Fermionic Independent Set and Laplacian of an independence complex are QMA-hard}
\author{Chaithanya Rayudu}
\affil{\small Department of Physics and Astronomy and Center for Quantum Information and Control,\\ University of New Mexico, Albuquerque, NM, USA}
\date{}

\begin{document}

\maketitle

\begin{abstract}
    The Independent Set is a well known NP-hard optimization problem. In this work, we define a fermionic generalization of the Independent Set problem and prove that the optimization problem is QMA-hard in a $k$-particle subspace using perturbative gadgets. We discuss how the \emph{Fermionic Independent Set} is related to the problem of computing the minimum eigenvalue of the $k^{\text{th}}$-Laplacian of an independence complex of a vertex weighted graph. Consequently, we use the same perturbative gadget to prove QMA-hardness of the later problem resolving an open conjecture from \cite{king2023promise} and give the first example of a natural topological data analysis problem that is QMA-hard.
   
    
\end{abstract}

\section{Introduction}

\onote{I'm guessing you'll replace this intro with the approach we took for the QIP abstract.  I would make sure to emphasize the following:
\begin{itemize}
    \item TDA has been studied classically for decades and has many applications
    \item The complexity of natural TDA was not understood and a resolution of this was an open problem
    \item Recent breakthrough work gave a good reason for why the complexity had been open: namely, these problems were quantum mechanical in disguise.
    \item However, the question of QMA-completeness of a natural TDA problem remained open, and we resolve it.
    \item Our proof approach is based on recent work in generalizing discrete optimization problems to quantum local Hamiltonians, in particular Vertex Cover/Independent Set.  This results in a substantially streamlined proof than previous approaches.
    \item Understanding classical problems from a quantum perspective is important for potentially new kinds of quantum advantages and discovering new BQP-complete problems. You may want to point out the contrast between QLSP and LSP here.  The former is a quantum version of LSP, and we know there is some advantage if quantum has any advantage at all (since it is BQP-complete).
\end{itemize}
I know you already have most of this below, but in case you want to add or rearrange things, I'll be happy to take another look after you've edited.  Maybe the main thing missing is setting up the story with the open-complexity-of-TDA-problems angle and mentioning connections to quantum advantages.}

Quantum Merlin Arthur (QMA) is the complexity class of decision problems for which a `yes' solution can be efficiently verified using a quantum computer. The $k$-Local Hamiltonian problem \cite{kitaev2002classical, kempe20033, kempe2005complexity} was the first non-trivial problem shown to be QMA-complete. Since then numerous other problems have been proven QMA-complete, mostly arising in quantum computing and information science \cite{bookatz2012qma}. 
While the $k$-Local Hamiltonian problem serves as a fundamental QMA-complete problem, analogous to $k$-SAT's role in NP-completeness, the majority of NP-complete problems are constrained combinatorial optimization problems such as Minimum Vertex Cover, Maximum Independent Set, and Longest Path \cite{vazirani2001approximation}. 
Such constrained optimization problems had a great impact on the development of classical algorithms, and hardness results.
By studying quantum generalizations of such constrained optimization problems, we can explore new quantum algorithms and hardness results for quantum problems that are currently lacking. 
This kind of motivation inspired a previous work \cite{parekh2024constrained} which studies quantum generalizations of classical constrained problems with Vertex Cover as a driving example.
Following this motivation, we define a fermionic generalization of the Independent Set and study its computational complexity. 
We prove that the Fermionic Independent Set is QMA-hard when restricted to a $k$-particle subspace.

When problems without an obvious quantum connection are shown to be related QMA, it can offer new insights into the structure of QMA and as well as the problem at hand. One such problem in Homology is to determine the existence of a $k$-dimensional hole in a topological space.
When the space is represented by the clique complex of a graph, this problem is of practical interest in Topological Data Analysis (TDA) which uses topological features such as holes to characterize and study data. 
TDA has found a broad range of applications from computational neuroscience \cite{sizemore2019importance} to cosmology \cite{heydenreich2021persistent}.
Recently, there has been a significant progress towards understanding the computational complexity of determining the existence of a hole by relating it to quantum complexity classes.
This connection was possible because of a correspondence between the combinatorial Laplacian of a clique complex whose null space captures the holes and a supersymmetric Hamiltonian \cite{cade2024complexity}.
The question of whether the clique complex has a $k$-dimensional hole or not is equivalent to whether the supersymmetric Hamiltonian has a zero energy ground state or not in the $k$-particle subspace. In this work, we prove that computing the minimum eigenvalue of the $k^\text{th}$-Laplacian of a clique complex is QMA-hard resolving an open conjecture from \cite{king2023promise} in the affirmative. Our result is proved for the Laplacian operator of the independence complex and holds true for the clique complex as well since the clique complex of a graph is same as the independence complex of the complement graph. A novelty of our work lies in the simplicity of our proof techniques based on perturbative gadgets compared to highly specialized homological proof techniques of previous works \cite{crichigno2022clique, king2023promise}.

Although the Fermionic Independent Set and the minimum eigenvalue problem of the Laplacian of an independence complex might seem unrelated at first, when expressed as constrained optimization problems they differ only slightly in their objective functions. The similarity between the problems is also reflected in the fact that we use the same perturbative gadget to prove QMA-hardness of both the problems. We consequently establish the first QMA-hardness result of a natural TDA problem.
\onote{As in the abstract, end with the punchline, ``We consequently establish the first QMA-completeness result...''}

\subsection{Related Work}
Crichigno and Kohler \cite{crichigno2022clique} proved that the problem of deciding whether the minimum eigenvalue of the $k^{\text{th}}$-Lapalacian is equal to $0$ or $>0$ is QMA$_1$-hard where QMA$_1$ is the complexity class QMA with perfect completeness.
But the corresponding QMA$_1$-hard supersymmetric Hamiltonians that they construct can have exponentially small gap above the zero energy ground space.
This was later improved by King and Kohler \cite{king2023promise} where they constructed QMA$_1$-hard supersymmetric Hamiltonians corresponding to clique complexes of vertex weighted graphs with inverse polynomial gap above the zero energy ground space, thereby placing them in QMA.
The authors conjectured in \cite{king2023promise} that there are supersymmetric Hamiltonians corresponding to clique complexes which are QMA-hard, meaning estimating the minimum eigenvalue of the Laplacian operator even when it is nonzero is QMA-hard, rather than just QMA$_1$-hard.

\subsection{Notation and organization}


The notation $[n]$ denotes the set of integers $\{1,2,...,n\}$. A graph is denoted by $G$, $G([n],E)$ where vertices are labeled by elements from $[n]$ and set $E$ denotes the edges between the vertices. An edge between vertices $i$ and $j$ is represent by $ij$.  Hamiltonians are denoted by $H$ and $V$ with various subscripts, and projectors are denoted by $P$ and $\Pi$ with various subscripts. Hilbert spaces are denoted by $\mathcal{H}$ with various subscripts. The notation $\rho$ denotes a density operator and the notation $\tr[A]$ denotes the matrix trace of an operator $A$. Positive semi-definiteness of an operator $A$ is denoted by $A\semigeq 0$. Fermionic creation and annihilation operators are denoted by  $a^{\dagger}$, $a$. Pauli matrices $X$, $Y$, and $Z$ are defined as follows
\begin{align*}
X=\begin{bmatrix}
0 & 1 \\
1 & 0
\end{bmatrix},
\,\,\,\,\,\,
Y=\begin{bmatrix}
0 & -i \\
i & 0
\end{bmatrix}, \,\text{and}
\,\,\,\,\,\,
Z=\begin{bmatrix}
1 & 0 \\
0 & -1
\end{bmatrix}.
\end{align*}

In \cref{sec:definitions}, we define the Fermionic Independent Set problem and the minimum eigenvalue problem of the Laplacian of an independence complex of a vertex weighted graph. We briefly the similarities and the differences between the two problems. In \cref{sec:perturbative_gadgets}, we recall some theorems behind perturbative gadgets necessary for our work. In \cref{sec:complexity_of_fermionic_independent_set}, we give a proof of QMA-hardness of Fermionic Independent Set problem when restricted to a $k$-particle subspace. In \cref{sec:complexity_of_laplacian_independence_complex}, we extend the QMA-hardness proof to minimum eigenvalue problem of the $k^{\text{th}}$-Laplacian of an independence complex.

\section{Definitions}
\label{sec:definitions}

\subsection{Independent Set}
Independent Set is a combinatorial graph optimization problem. Given a graph $G([n], E)$ with vertex weights $\{c_i\}_{i\in [n]}$, a subset of vertices $S \subseteq [n]$ is called an independent set of graph $G$ if no two vertices from the subset $S$ are adjacent to each other in graph $G$. In other words, $S$ is an independent set of $G$ if for all distinct $i,j \in S$, $ij \notin E$. Weight of an independent set $S$ is sum of weights of all the vertices in the independent set $S$. The weighted Independent Set problem is to find an independent set with the maximum weight. Let $x_i \in \{0,1\}$ be a binary variable for each $i \in [n]$ such that if $i \in S$ then $x_i = 1$ and $x_i = 0$ otherwise. Formally, we can define the optimization problem as
\begin{align}\label{eq:IS-integer-program}
    &\max\,\, \sum_{i\in V} c_i x_i\\
     \label{cons:IS}
     s.t.\,\,\,\,\,\,\,\,\,\,\, & x_i x_j =0 \quad \forall\, ij \in E,\\  
    & x_i \in \{0,1\} \quad \forall\, i \in [n].
\end{align}
The constraint $x_ix_j=0$ implies at most only one of $x_i$ or $x_j$ can be equal $1$ which implies at most only one of $i$ or $j$ can be in set $S$. Independent Set and Vertex Cover are complement to each other. For every independent set $S \subseteq [n]$, the complement subset $[n]\backslash S$ is a vertex cover. Therefore, Maximum Independent Set and Minimum Vertex Cover problems are equivalent to each other.

By identifying the vertices of the graph with fermionic modes such that creation and annihilation operators of the mode labeled by the $i^\text{th}$ vertex are $a_i^{\dagger}$ and $a_i$, we can represent an independent set $S$ with fock states $\Pi_{i\in S}\, a^{\dagger}_i\ket{\text{vac}}$ that are equivalent up to a phase. With the fermionic representation, we can rewrite the Independent Set optimization problem as
\begin{align}
    \max_{\rho}\,\, &\tr\left[\left(\sum_ic_ia^{\dagger}_ia_i\right)\rho\right]\\
    s.t.\quad &\tr\left[a^{\dagger}_ia_ia^{\dagger}_ja_j\rho\right] = 0\quad \forall\, ij \in E,\label{eq:independent_set_constraint}\\
    &\tr[\rho] = 1,\,\, \rho \semigeq 0.
\end{align}
The constraint operators $a^{\dagger}_ia_ia^{\dagger}_ja_j$ in \cref{eq:independent_set_constraint} are projectors onto the fock state $a_i^{\dagger} a_j^{\dagger}\ket{vac}$ of the $i$,$j$ modes restricting $\rho$ to be in the subspace orthogonal to $a_i^{\dagger} a_j^{\dagger}\ket{vac}$. 
These constraints require $\rho$ to have a maximum of one spinless fermion among the modes $i$ and $j$ for every edge $ij$ in graph $G$. 

\subsection{Fermionic Independent Set}

\onote{For any problem you solve, you should never let the reader ask themselves the question: why should I care?  They should have already been given a clear answer.  So natural questions that could come up here are: while I know people care about fermionic Hamiltonians, why is this Hamiltonian one I should care about?  Yes, it helps resolve QMA-completeness of a TDA problem, but does it connect to other Hamiltonians people already care about?  Does it add any additional value to those?  In this case, you explain things for someone looking to generalize IS, but for someone who doesn't care so much about generalizing IS, why should they still care about this Hamiltonian?  I think you have implicit answers below, but maybe explicitly state them as ``features'' that add new elements to previously studied Hamiltonians.}

We propose a fermionic generalization of the Independent Set called \emph{Fermionic Independent Set} defined as follows.

\begin{definition}[Fermionic Independent Set]
\label{def:fermionic_independent_set}
Let $G([n], E)$ be a graph with $n$ vertices and an edge set $E$ where the vertices label fermionic modes that can be occupied by spinless fermions. Given a hopping Hamiltonian $H = \sum_{ij \in E} w_{ij} (a^{\dagger}_i a_j + a^{\dagger}_j a_i)$, the Fermionic Independent Set optimization problem is defined as
    \begin{align}
        \min_{\rho} \,\, &\tr[ \left(\sum_{ij \in E} w_{ij} (a^{\dagger}_i a_j + a^{\dagger}_j a_i)\right) \rho]\\
        s.t.\quad &\tr[a^{\dagger}_ia_ia^{\dagger}_ja_j\rho] = 0 \quad \forall\, ij \in E, \label{eq:independent_set_constraint_2}\\
        &\tr[\rho] = 1,\,\, \rho \semigeq 0.
    \end{align}
\end{definition}

One of the key features we aimed for in defining a fermionic generalization of the classical Independent Set is that the objective Hamiltonian should be easy to optimize if we ignore the constraints. With that in mind, our choice of Hamiltonian $H$ to be a hopping Hamiltonian is not the only choice.
One could consider a more general quadratic Hamiltonian that includes diagonal terms $a^{\dagger}_i a_i$ which can be exactly diagonalized in a polynomial time on a classical computer.
Without the quadratic diagonal terms, our definition of a Fermionic Independent Set is not a strict generalization of the classical Independent Set problem. But for the purpose of understanding the computational complexity of the problem, it is sufficient to restrict to single particle hopping Hamiltonians. 

Instead of fermionic generalization of the Independent Set, we could consider generalizing the Vertex Cover 
with fermions where we have vertex cover constraints instead of independent set constraints. Physically, independent set constraints can be intuitively interpreted as fermionic modes with strong repulsion giving rise to an effective low energy subspace.
One way to obtain the fermionic generalization of Vertex Cover from \cref{def:fermionic_independent_set} is by applying a transformation $\mathcal{T}$ that transforms the creation and annihilation operators as $\mathcal{T}a_i\mathcal{T}^{-1} = a_i^{\dagger}$, $\mathcal{T}a_i^{\dagger}\mathcal{T}^{-1} = a_i$ and the vacuum state as $\mathcal{T}\ket{vac} = \Pi_{i=1}^{|V|}a_i^{\dagger}\ket{vac}$. 
This transformation changes \cref{eq:independent_set_constraint_2} into a vertex cover constraint $\tr[(I-a^{\dagger}_ia_i)(I-a^{\dagger}_ja_j)\rho] = 0$ and the hopping Hamiltonian from $H$ to $-H$. 
Therefore, both the problems are equivalent to each other.


In \cref{sec:complexity_of_fermionic_independent_set}, we prove that the Fermionic Independent Set is QMA-hard when restricted to a $k$-particle subspace. 
\begin{restatable}{theorem}{thmFIS}
\label{thm:FIS_QMAhardness}
    The Fermionic Independent Set problem as defined in \cref{def:fermionic_independent_set} is QMA-hard when restricted to a $k$-particle subspace.
\end{restatable}
Furthermore, we prove QMA-hardness of the problem with additional structure thereby implying that the general problem is also QMA-hard. The additional structure is that the hopping weights $w_{ij} = u_i u_j$ $\forall\, ij \in E$ with vertex weights $u_i \geq 0 \,\, \forall\, i \in [n]$. 

\subsection{Laplacian of an independence complex}
\label{intro:independence_complex}

\onote{You want people to read and refer to your paper.  If the reader gets confused and has to look somewhere else, they may not come back.  For such reasons, if you can add any context or introductory material on TDA that would help set the stage, it would be useful.  Even a paragraph or two with good and easy to read references could help.

In the case of TDA, people are in general curious about it and will also look at your paper because the result seems more technically accessible than previous approaches.  Make sure your paper accessible overall.  Like I said, you should aim to provide a complete story (but not necessary including all details) of QMA-completeness of TDA problems.}

In topological data analysis, data is studied using topological features such as holes. When the data is given as a graph which in turn can be triangulation of an underlying topological manifold, homology of the clique complex of the graph provides information about the holes in the underlying topological manifold. Clique complex of a graph is equivalent to independence complex of the complement graph.

The independence complex of a graph $G([n],V)$ is a simplical chain complex where the simplices are the independent sets of the graph $G$. Let $S = \{v_1,v_2,...,v_k\} \subseteq [n]$ be an independent set of size $k$ of the graph $G$, and the corresponding $k$-simplex is an ordered set of vertices $\ket{v_1,v_2,...,v_k}$. The simplices that differ only in the ordering of vertices are equivalent to each other up to a phase such that
\begin{align}
    |...,v_i,v_j,...\rangle = -|...,v_j,v_i,...\rangle
\end{align}
where swapping of two neighboring vertices in the ordering induces a negative sign. Let $C_k$ be the free abelian group with $k$-simplices as the basis. The boundary operator $\partial_k: C_k \rightarrow C_{k-1}$ is defined by its action on a $k$-simplex as
\begin{align}
    \partial_k\ket{v_1,...,v_k} = \sum_{i=1}^k (-1)^i \ket{v_1,...v_{i-1},v_{i+1},...,v_k},
\end{align}
and the boundary operators satisfy $\partial_k\partial_{k-1} = 0\, \forall\, k$. Formally, independence complex is a chain complex defined as
\begin{align*}
    ... \xrightarrow{\partial_{k+2}} C_{k+1} \xrightarrow{\partial_{k+1}} C_k \xrightarrow{\partial_{k}} C_{k-1} \xrightarrow{\partial_{k-1}} ...
\end{align*}
The $k^{\text{th}}$-homology $\mathcal{H}_k$ of the chain complex is the following quotient group:
\begin{align}
    \mathcal{H}_k = \text{Kernel}(\partial_k)/\text{Image}(\partial_{k+1}).
\end{align}
We can capture the $k^{\text{th}}$-Homology group using representatives from each equivalence class in the quotient group that are orthogonal to Image$(\partial_{k+1})$. To have a notion of orthogonality, let us introduce an inner product $\langle a,b \rangle$ between any two simplices $a,b$ as follows
\begin{align}
   \langle a,b\rangle =
   \begin{cases}
       w(a)w(b) & \text{ if }a=b\\
       -w(a)w(b) & \text{ else if }a=-b\\
       0 & \text{ otherwise}
   \end{cases}
\end{align}
where weight of a simplex $w(v_1,v_2,...,v_k) = w(v_1)w(v_2)...w(v_k)$ is the product of weights of its vertices. With this inner-product, we now have a Hilbert space. The span of the representatives from each of the equivalence classes is the kernel of the $k^\text{th}$-Laplacian operator $\Delta_k: C_k \rightarrow C_k$ defined as
\begin{align}
    \Delta_k = \partial_k^{\dagger}\partial_k + \partial_{k+1}\partial_{k+1}^{\dagger}.
\end{align}
Note that the Laplacian operator is positive semi-definite. The question of whether there are any $k$-dimensional holes is equivalent to asking if the minimum eigenvalue of the $k^\text{th}$-Laplacian $\Delta_k$ is $=0$ or $>0$.

We can rewrite the question about minimum eigenvalue of the Laplacian using fermions. Let the vertices of the graph $G([n],E)$ label fermionic modes that can be occupied by spinless fermions, and consider a staggered (weighted) supersymmetric Hamiltonian $H_{\text{Lap}} = Q^\dagger Q + Q Q^\dagger$, where 
\begin{align}
    Q=\sum_{i \in [n]} u_i P_i a_i \quad \text { and } \quad Q^{\dagger}=\sum_{i \in [n]} u_i  P_ia_i^{\dagger} \quad \text { where } \quad P_i=\prod_{j \mid ij \in E}\left(I-a^{\dagger}_ja_j\right) \quad \text { and } \quad u_i \geq 0
\end{align}
After simplification, we can rewrite the Hamiltonian $H_{\text{Lap}}$ as
\begin{align}
    \label{eq:susyH}
    H_{\text{Lap}} = \sum_{ij\in E} u_iu_j P_ia_i^{\dagger}a_jP_j + \sum_{i \in [n]} u_i^2 P_i.
\end{align}
It is easy to verify that the Hamiltonian $H_{\text{Lap}}$ preserves the subspace of hard-core fermions defined as the fock subspace where no two fermions are adjacent to each other on the graph $G$ which is an independent constraint. The subspace of hard-core fermions forms a representation for the simplices of the independence complex, and the Hamiltonian $H_{\text{Lap}}$ projected onto the subspace of hardcore fermions forms a representation for the Laplacian of the independence complex \cite{crichigno2022clique, huijse2010supersymmetry}. Now the question about the minimum eigenvalue of the Laplacian can be rewritten as a constrained optimization problem using fermions as follows.

\begin{definition}[Minimum eigenvalue of the Laplacian of an independence complex]
    Let $G([n], E)$ be a graph with $n$ vertices and an edge set $E$ where the vertices label fermionic modes that can be occupied by spinless fermions. Let $H_\text{Lap}$ be a supersymmetric Hamiltonian defined as
    \begin{align}
        H_\text{Lap} = \sum_{ij\in E} u_iu_j (a_i^{\dagger}a_j + a_j^{\dagger}a_i) + \sum_{i \in V} u_i^2 P_i \quad \text{where}\quad P_i=\prod_{j \mid ij \in E}\left(I-a^{\dagger}_ja_j\right) \quad \text { and } \quad u_i \geq 0.
    \end{align}
    Then the problem of finding the minimum eigenvalue of $H_{\text{Lap}}$ in the subspace of hard-core fermions can be written as a constrained optimization problem as follows
    \begin{align}
        \min_{\rho} \,\, &\tr\left[\left(\sum_{ij\in E} u_iu_j (a_i^{\dagger}a_j + a_j^{\dagger}a_i) + \sum_{i \in [n]} u_i^2 P_i\right)\rho\right]\\
        s.t.\quad &\tr[a^{\dagger}_ia_i a^{\dagger}_ja_j \rho] = 0 \quad \forall\, ij \in E,\\
        &\tr[\rho] = 1,\,\, \rho \semigeq 0.
    \end{align}
\end{definition}

There are two main differences between the above optimization problem and the definition \ref{def:fermionic_independent_set} of Fermionic Independent Set. First, the weights for the hopping terms $(a_i^{\dagger}a_j + a_j^{\dagger}a_i)$ in the objective have a vertex weighted product structure for edge weights i.e., $w_{ij} = u_i u_j\,\forall ij \in E$. Second, the objective Hamiltonian has additional diagonal terms $\sum_{i \in V} u_i^2 P_i$.

It was shown in \cite{king2023promise} that the promise problem of deciding whether the optimal value to the above problem is zero or at least $1/\text{poly}(n)$ in a given $k$-particle subspace is QMA$_1$-hard and is contained in QMA, and it was conjectured that the promise problem of deciding whether the optimal value is $\leq a$ ($a$ need not be equal to zero) or $\geq b$ for some $a,b$ such that $b-a \geq 1/\text{poly}(n)$ is QMA-hard. We resolve this conjecture in \cref{sec:complexity_of_laplacian_independence_complex} by showing that it is indeed QMA-hard to estimate the optimal value within an inverse polynomial additive precision even when it is nonzero.

\section{Complexity of Fermionic Independent Set}
\label{sec:complexity_of_fermionic_independent_set}


\thmFIS*

We show this by a reduction from $XZ$-Hamiltonian which was proven to be QMA-hard and universal for quantum simulation in \cite{Piddock2015TheCO, cubitt2018universal}. 
The rest of this section contains the proof of \cref{thm:FIS_QMAhardness}. 
An overview of the perturbative gadget in the proof is as follows.
\begin{enumerate}
    \item We start by encoding a qubit using three fermionic modes. The three modes have a pairwise independent set constraint between them as defined in \cref{eq:independent_set_constraint} and these constraints define the accessible subspace to be the span of single particle subspace and vacuum. In the accessible subspace, we encode a qubit as the low energy two-dimensional ground subspace of a single particle hopping Hamiltonian.

    \item Next, we want to simulate the $XZ$ interaction between any pair of encoded qubits. For every pair of encoded qubits between which we want to create $XZ$-interaction, we add three additional mediator modes. We also add additional independent set constraints as defined by the edges of the graph in \cref{fig:interaction_constraint_graph} and a perturbative single particle hopping Hamiltonian to create $XZ$-interaction at second order perturbation. 
\end{enumerate}

This construction guarantees that the ground state energy of any given target $XZ$-Hamiltonian can be made arbitrarily close to the optimal value of the Fermionic Independent Set restricted to a $k$-particle subspace. Since estimating the ground state energy of a $XZ$-Hamiltonian is QMA-hard, Fermionic Independent Set when restricted to a $k$-particle is also QMA-hard.
A key point in our analysis is that we do all our calculations in the accessible subspace defined by the independent set constraints.

\subsection{Encoding a qubit} \label{sec:encoding_a_qubit}

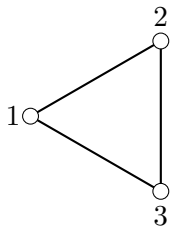
\begin{figure}[h!]
    \centering
    \begin{tikzpicture}
        \draw[black, thick] (-1.732,0) -- (0,1);
        \draw[black, thick] (-1.732,0) -- (0,-1);
        \draw[black, thick] (0,-1) -- (0,1);
        \draw[fill=white] (-1.732,0) circle [radius=3pt];
        \draw[fill=white] (0,1) circle [radius=3pt];
        \draw[fill=white] (0,-1) circle [radius=3pt];
        \node[left] at (-1.732,0) {$1$};
        \node[above=2pt] at (0,1) {$2$};
        \node[below=2pt] at (0,-1) {$3$};
    \end{tikzpicture}
    \caption{Independent set constraint graph of three fermionic modes labeled 1,2,3.}
    \label{fig:triangle_constraint_graph}
\end{figure}

Consider three fermionic modes labeled ${1,2,3}$, a constraint graph with edges $(1,2)$, $(2,3)$ and $(1,3)$ as in \cref{fig:triangle_constraint_graph} which requires at least two spinless fermions to satisfy the independent set constraints,
and consider a hopping Hamiltonian 
\begin{align}
    \label{eq:triangle_hopping_hamiltonian}
    H_0 = I+(a_1^{\dagger}a_2+a_2^{\dagger}a_1) +(a_2^{\dagger}a_3+a_3^{\dagger}a_2) +(a_3^{\dagger}a_1+a_1^{\dagger}a_3)
\end{align}
on the three modes which block diagonalizes into one, two, and three particle sectors. Among the eigenstates of $H_0$ that satisfy the independent set constraints, there are two $1$-particle states with energy equal to $0$, one $1$-particle state with energy equal to $3$, and the vacuum with energy equal to $1$. Following are the corresponding eigenstates of $H_0$
\begin{align}
    \ket{0} \,\,&:=\,\, S_{0} \ket{vac},\quad \ket{1} \,\,:=\,\,  S_{1} \ket{vac},\label{eq:encoded_qubit}\\
    \ket{2} \,\,&:=\,\,  S_{2} \ket{vac},\quad \ket{3} \,\,:=\,\,  \ket{vac}
\end{align}
where $\ket{vac}$ is the vacuum state and
\begin{align}
    S_{0} := \frac{1}{2\sqrt{3}}\left[(1-\sqrt{3})a_3^{\dagger} - 2a_1^{\dagger} + (1 +\sqrt{3})a_2^{\dagger}\right], \quad
    &S_{1} := \frac{1}{2\sqrt{3}}\left[(1+\sqrt{3})a_3^{\dagger} - 2a_1^{\dagger} + (1-\sqrt{3})a_2^{\dagger}\right],\label{eq:encoded_01_operators}\\
    S_{2}  := \frac{1}{\sqrt{3}}&\left[a_3^{\dagger} + a_1^{\dagger} + a_2^{\dagger}\right]
\end{align}
The states $\ket{0}$ and $\ket{1}$ have energy equal to 0 while the states $\ket{2}$ and $\ket{3}$ have energy equal to 3 and 1 respectively. We can now encode a qubit into the 2-dimensional low energy subspace spanned by the states $\ket{0}$ and $\ket{1}$. 

\subsection{\texorpdfstring{$XZ$}{XZ}-interaction between encoded qubits}

\begin{figure}[h!]
    \centering
    \begin{tikzpicture}
        \draw[black, thick] (-1.732-0.5,0) -- (0-3.5,1);
        \draw[black, thick] (-1.732-0.5,0) -- (0-3.5,-1);
        \draw[black, thick] (0-3.5,-1) -- (0-3.5,1);

        \draw[black, thick] (+1.732+0.5,0) -- (0+3.5,1);
        \draw[black, thick] (+1.732+0.5,0) -- (0+3.5,-1);
        \draw[black, thick] (0+3.5,-1) -- (0+3.5,1);

        \draw[black, thick] (0-3.5,1) -- (0,1) -- (0+3.5,1);
        \draw[black, thick] (0-3.5,-1) -- (0,-1) -- (0+3.5,-1);
        \draw[black, thick] (-1.732-0.5,0) -- (0,0) -- (+1.732+0.5,0);

        \draw[fill=white] (+1.732+0.5,0) circle [radius=3pt];
        \draw[fill=white] (0+3.5,1) circle [radius=3pt];
        \draw[fill=white] (0+3.5,-1) circle [radius=3pt];

        \draw[fill=white] (-1.732-0.5,0) circle [radius=3pt];
        \draw[fill=white] (0-3.5,1) circle [radius=3pt];
        \draw[fill=white] (0-3.5,-1) circle [radius=3pt];

        \draw[fill=white] (0,1) circle [radius=3pt];
        \draw[fill=white] (0,-1) circle [radius=3pt];
        \draw[fill=white] (0,0) circle [radius=3pt];

        \node[above right] at (-1.732-0.5,0) {$i1$};
        \node[above=2pt] at (0-3.5,1) {$i2$};
        \node[below=2pt] at (0-3.5,-1) {$i3$};
        
        \node[above left] at (+1.732+0.5,0) {$j1$};
        \node[above=2pt] at (0+3.5,1) {$j2$};
        \node[below=2pt] at (0+3.5,-1) {$j3$};

        \node[above=2pt] at (0,1) {$x$};
        \node[below=2pt] at (0,-1) {$y$};
        \node[above=2pt] at (0,0) {$z$};
        
    \end{tikzpicture}
    \caption{Independent set constraint graph to create interaction between two encoded qubits $i$ and $j$.}
    \label{fig:interaction_constraint_graph}
\end{figure}
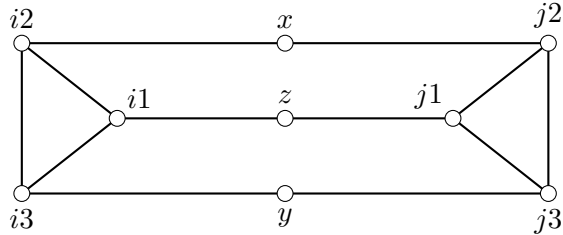

Consider sets of three fermionic modes labeled by $\{i1, i2, i3\}$ for every logical qubit-$i$ encoded using the isometry defined in the previous section. To simulate a two-qubit interaction between the qubits $i$ and $j$, we add three empty mediator modes labeled $x$,$y$,$z$ for every pair of $i,j$ qubits and modify the independent set constraint graph by adding additional edges $(i2,x)$, $(i3,y)$, $(j2,x)$, $(j3,y)$, $(i1,z)$ and $(j1,z)$ as depicted in \cref{fig:interaction_constraint_graph}.

We will consider the same hopping Hamiltonian $H_{0,i} + H_{0,j}$ corresponding to $\{i1,i2,i3\}$ and $\{j1,j2,j3\}$ modes as defined in \cref{eq:triangle_hopping_hamiltonian}. Within the subspace defined by the independent set constraints of the graph in \cref{fig:interaction_constraint_graph} with 2 particles, the low energy subspace of $H_{0,i} + H_{0,j}$ is $4$-fold degenerate with the following eigenstates
\begin{align}
    \label{eq:two_qubit_encoded_states}
    &\ket{00}_{ij} := S_{i,0} S_{j,0} \ket{vac},\quad \ket{01}_{ij} := S_{i,0} S_{j,1} \ket{vac}, \nonumber\\
    &\ket{10}_{ij} := S_{i,1} S_{j,0} \ket{vac},\quad \ket{11}_{ij} := S_{i,1} S_{j,1} \ket{vac}.
\end{align}
where $S_{i,0}, S_{i,1}, S_{j,0}$, $S_{j,1}$ as in \cref{eq:encoded_01_operators}. Let $\Pi_{-}$ be the projector onto the subspace spanned by these four encoded basis states and let $\Pi_{+}$ be the projector onto the rest of the Hilbert space satisfying the independent set constraints of the graph in \cref{fig:interaction_constraint_graph} with $2$ particles.

To simulate a two-qubit interaction between the encoded qubits $i$ and $j$, let us consider a perturbation 
\begin{align}
\label{eq:perturbation_for_FIS}
    V_{main} = V_1 + V_2
\end{align}
where
\begin{align}
     V_1 &= (a_{i2}^{\dagger}a_x + a_x^{\dagger}a_{i2})+(a_{j2}^{\dagger}a_x + a_x^{\dagger}a_{j2})+(a_{i3}^{\dagger}a_y + a_y^{\dagger}a_{i3})+(a_{j3}^{\dagger}a_y + a_y^{\dagger}a_{j3}). \\
     V_2 &= (a_{i1}^{\dagger}a_z + a_z^{\dagger}a_{i1})+(a_{j1}^{\dagger}a_z + a_z^{\dagger}a_{j1}).
\end{align}
To first-order perturbation, there is no effect on the encoded qubits since $(V_{main})_{--} = 0$. At second-order perturbation, $V_{main}$ can give rise to nontrivial interaction between the encoded qubits. To see the effect of $V_{main}$, let us first consider the effect of only a part of the perturbation $V_1^{'} = (a_{i2}^{\dagger}a_x + a_x^{\dagger}a_{i2})$ as follows
\begin{align}
    (V_1^{'})_{+-}\ket{00}_{ij} &= \frac{(1+\sqrt{3})}{12}a_{x}^{\dagger}\left[(1-\sqrt{3})a_{j3}^{\dagger} - 2a_{j1}^{\dagger}\right]\ket{vac}\\
    (V_1^{'})_{+-}\ket{01}_{ij} &= \frac{(1+\sqrt{3})}{12} a^{\dagger}_{x}\left[(1+\sqrt{3})a_{j3}^{\dagger} - 2 a_{j1}^{\dagger}\right]\ket{vac},\\
    (V_1^{'})_{+-}\ket{10}_{ij} &= \frac{(1-\sqrt{3})}{12} a^{\dagger}_{x}\left[(1-\sqrt{3})a_{j3}^{\dagger} - 2 a_{j1}^{\dagger}\right]\ket{vac},\\
    (V_1^{'})_{+-}\ket{11}_{ij} &= \frac{(1-\sqrt{3})}{12} a^{\dagger}_{x}\left[(1+\sqrt{3})a_{j3}^{\dagger} + 2 a_{j1}^{\dagger}\right]\ket{vac}.
\end{align}
The effect of operator $(H_{0,i} + H_{0,j})^{-1}$ acting on above states can be calculated by writing them as a superposition of the following excited eigenstates of $(H_{0,i} + H_{0,j})$,
\begin{align}
    \frac{1}{\sqrt{2}}a^{\dagger}_{x}\left(a^{\dagger}_{j1} + a^{\dagger}_{j3}\right)\ket{vac} \quad \text{ and } \quad \frac{1}{\sqrt{2}}a^{\dagger}_{x}\left(a^{\dagger}_{j1} - a^{\dagger}_{j3}\right)\ket{vac}
\end{align}
with eigenvalues $3$ and $1$ respectively. For example,
\begin{align}
    (H_{0,i} +& H_{0,j})^{-1}(V_1^{'})_{+-} \ket{00}_{ij} \nonumber\\
    &= (H_{0,i} + H_{0,j})^{-1} \frac{(1+\sqrt{3})}{12} a^{\dagger}_{x}\left[\frac{(3-\sqrt{3})}{\sqrt{2}}\frac{\left(-a^{\dagger}_{j1} + a^{\dagger}_{j3}\right)}{\sqrt{2}} + \frac{(-1-\sqrt{3})}{\sqrt{2}}\frac{\left(a^{\dagger}_{j1} + a^{\dagger}_{j3}\right)}{\sqrt{2}}\right]\ket{vac}\\
    &= \frac{(1+\sqrt{3})}{12} a^{\dagger}_{x}\left[\frac{(3-\sqrt{3})}{\sqrt{2}}\frac{\left(-a^{\dagger}_{j1} + a^{\dagger}_{j3}\right)}{\sqrt{2}} + \frac{(-1-\sqrt{3})}{3\sqrt{2}}\frac{\left(a^{\dagger}_{j1} + a^{\dagger}_{j3}\right)}{\sqrt{2}}\right]\ket{vac}
\end{align}
After some more similar calculations, we can express the effective interaction $(V_1^{'})_{-+}(H_{0,i} + H_{0,j})^{-1}(V_1^{'})_{+-}$ in the matrix form as follows
\begin{align}
    -(V_1^{'})_{-+}(H_{0,i} + H_{0,j})^{-1}(V_1^{'})_{+-} = \frac{-1}{54}
    \begin{pmatrix}
        4+\sqrt{3} & 4+2\sqrt{3} & -5+2\sqrt{3} & -2\\
        4+2\sqrt{3} & 16+9\sqrt{3} & -2 & -5-2\sqrt{3}\\
        -5+2\sqrt{3} & -2 & 16-9\sqrt{3} & 4-2\sqrt{3}\\
        -2 & -5-2\sqrt{3} & 4-2\sqrt{3} & 4-\sqrt{3}
    \end{pmatrix}
\end{align}
where the rows and columns are index by $\left\{\ket{00},\ket{01},\ket{10},\ket{11}\right\}_{ij}$ basis states defined in \cref{eq:two_qubit_encoded_states}. Symmetrizing the interaction $V_1^{'}$ with respect to $X_iX_j$ and $SWAP(i,j)$ operators gives the effective interaction due to perturbation $V_1$ as follows
\begin{align}
    -(V_1)_{-+}(H_{0,i} + H_{0,j})^{-1}(V_1)_{+-} &= \frac{-1}{27}
    \begin{pmatrix}
        8 & -1 & -1 & -4\\
        -1 & 32 & -4 & -1\\
        -1 & -4 & 32 & -1\\
        -4 & -1 & -1 & 32
    \end{pmatrix}.\\
    &= \frac{1}{27}\left(-20I+4X_iX_j+12Z_iZ_j+X_i+X_j\right).
\end{align}
The second equality in the above equation is a rewriting in the Pauli basis of the encoded qubits. A similar calculation for $V_2$ shows that
\begin{align}
    -(V_2)_{-+}(H_{0,i} + H_{0,j})^{-1}(V_2)_{+-} = \frac{1}{27}\left(-10I+8X_iX_j-X_i-X_j\right).
\end{align}
Since $V_1$ perturbation uses $x,y$ as mediator modes and $V_2$ perturbation uses $z$ mediator mode, they can be applied in parallel to get
\begin{align}
    -(V_{main})_{-+}(H_{0,i} + H_{0,j})^{-1}(V_{main})_{+-} = -\frac{10}{9}I+\frac{4}{9}(X_iX_j+Z_iZ_j).
\end{align}

By \cref{thm:second_order_simulation} with $V_{extra} = 0$, we can simulate the spectrum of a target $XZ$-Hamiltonian $H_{target} = \sum_{ij} \frac{4}{9}\mu_{ij} (X_iX_j + Z_iZ_j)$ where $\mu_{ij} \geq 0 \forall i\neq j$ using Fermionic Independent Set with the objective Hamiltonian
\begin{align}
    H = \Delta \sum_{i} H_{0,i} + \Delta^{1/2}\sum_{ij} \sqrt{\mu_{ij}}\,(V_{main})_{ij}.
\end{align}
Since estimating the ground energy of $XZ$-Hamiltonian is QMA-hard, this completes the proof that Fermionic Independent Set is QMA-hard. A detail that will be important in the next section is that in the above Hamiltonian $H$ the weights of the hopping terms between different modes factorize product of single mode weights with the modes labeled by $\{i1,i2,i3\}\, \forall i$ having a weight $\Delta^{1/2}$ and the mediator modes between $\{i1,i2,i3\}$ and $\{j1,j2,j3\}$ having a weight $\sqrt{\mu_{ij}}$.

\section{Complexity of Laplacian of an independence complex}
\label{sec:complexity_of_laplacian_independence_complex}

The main difference between the Fermionic Independent Set with the vertex-weighted product structure for the edge weights and the minimum eigenvalue problem of the Laplacian of the independence complex of a vertex-weighted graph is the extra diagonal terms in the objective. To prove QMA-hardness of the later problem, we will use the same gadget we used to prove the QMA-hardness of Fermionic Independent Set but with additional diagonal terms in the objective. The analysis of the gadget is very similar in both cases. So the reader is highly recommended to go through the analysis of the gadget for the Fermionic Independent Set before going through rest of this section as we will only explain the differences in the gadget here.

Recall that in the gadget, every logical qubit $i$ from the target Hamiltonian is encoded using three physical modes $\{i1,i2,i3\}$ with a triangle graph as the independent set constraint graph. We modify the unperturbed Hamiltonian corresponding to each encoded qubit with diagonal terms compared to the case of the Fermionic Independent Set as follows
\begin{align}
\label{eq:unpertrubed_H_independence_complex}
    H_0 = (a_1^{\dagger}a_2+a_2^{\dagger}a_1) +(a_2^{\dagger}a_3+a_3^{\dagger}a_2) +(a_3^{\dagger}a_1+a_1^{\dagger}a_3)+ P_1+P_2+P_3
\end{align}
where
\begin{align}
    P_1 = (I-a_2^{\dagger}a_2)(I-a_3^{\dagger}a_3)...,\quad
    P_2 = (I-a_1^{\dagger}a_1)(I-a_3^{\dagger}a_3)...,\quad
    P_3 = (I-a_1^{\dagger}a_1)(I-a_2^{\dagger}a_2)...
\end{align}
The dots indicate other terms for mediator modes that we add to the gadget construction. For example, if we have two logical qubits $i,j$ being encoded using six physical modes $\{i1,i2,i3\}$, $\{j1,j2,j3\}$, and we use an additional three mediator modes $\{x,y,z\}$ to mediate the interaction between the encoded qubits $i,j$ with a constrained graph in \cref{fig:interaction_constraint_graph} in \cref{sec:complexity_of_fermionic_independent_set}, then
\begin{align}
    P_{i1} = (I-a_{i2}^{\dagger}a_{i2})(I-a_{i3}^{\dagger}a_{i3})&(I-a_z^{\dagger}a_z),\quad P_{i2} = (I-a_{i1}^{\dagger}a_{i1})(I-a_{i3}^{\dagger}a_{i3})(I-a_x^{\dagger}a_x),\nonumber\\
    P_{i3} &= (I-a_{i1}^{\dagger}a_{i1})(I-a_{i2}^{\dagger}a_{i2})(I-a_y^{\dagger}a_y)
\end{align}
and similarly for encoded qubit $j$. In the subspace with no fermions in the mediator modes, the spectrum of eigenstates of $H_0$ in \cref{eq:unpertrubed_H_independence_complex} that satisfy the independent set constraints remains the same as the unperturbed Hamiltonian in \cref{eq:triangle_hopping_hamiltonian}. Therefore, we will use the same isometry for encoding logical qubits as in the case of the Fermionic Independent Set.

To create interaction between the two encoded qubits $i$ and $j$, we will consider the same perturbation $V_{main} = V_1 + V_2$ as defined in \cref{eq:perturbation_for_FIS} via mediator modes. To see the effect of perturbation $V_{main}$ at second order, we can go through a similar analysis as in the case of the Fermionic Independent Set. One difference is the action of $(H_{0,i}+H_{0,j})^{-1}$ on the states $(V_{main})_{+-}\{\ket{00},\ket{01},\ket{10},\ket{11}\}$ because of the change of the eigenvalues for the following eigenstates of $(H_{0,i}+H_{0,j})$
\begin{align}
    \frac{1}{\sqrt{2}}a^{\dagger}_{x}\left(a^{\dagger}_{j1} + a^{\dagger}_{j3}\right)\ket{vac} \quad \text{ and } \quad \frac{1}{\sqrt{2}}a^{\dagger}_{x}\left(a^{\dagger}_{j1} - a^{\dagger}_{j3}\right)\ket{vac}
\end{align}
from 3 and 1 respectively to 4 and 2. Taking into account the differences in these eigenvalues, we get the following second-order perturbative effects due to $V_1$ and $V_2$
\begin{align}
    -(V_1)_{-+}(H_{0,i} + H_{0,j})^{-1}(V_1)_{+-} &= \frac{1}{72}\left(-28I+5X_iX_j+15Z_iZ_j+2X_i+2X_j\right),\\
    -(V_2)_{-+}(H_{0,i} + H_{0,j})^{-1}(V_2)_{+-} &= \frac{1}{72}\left(-14I+10X_iX_j-2X_i-
2X_j\right)
\end{align}
Combining the effects of $V_1$ and $V_2$, we get the effect due to $V_{main}$ as follows
\begin{align}
    -(V_{main})_{-+}(H_{0,i} + H_{0,j})^{-1}(V_{main})_{+-} = -\frac{7}{12}I+ \frac{5}{24}(X_iX_j+Z_iZ_j)
\end{align}

Unlike the case of the Fermionic Independent Set where we had $V_{extra} = 0$, we will consider the remaining diagonal terms corresponding to the mediator modes as $V_{extra}$. For example in the case of two encoded qubits $i,j$ with three mediator modes $\{x,y,z\}$,
\begin{align}
    V_{extra} = P_x + P_y + P_z
\end{align}
where
\begin{align}
    P_x = (I-a_{i2}^{\dagger}a_{i2})(I-a_{j2}^{\dagger}a_{j2}),\quad
    P_y = (I-a_{i3}^{\dagger}a_{i3})(I-a_{j3}^{\dagger}a_{j3}),\quad
    P_z = (I-a_{i1}^{\dagger}a_{i1})(I-a_{j1}^{\dagger}a_{j1}).
\end{align}
At first-order perturbation,
\begin{align}
    (V_{extra})_{--} = (P_x +P_y+P_z)_{--} = \frac{4}{3}I + \frac{1}{6}(X_iX_j + Z_iZ_j)
\end{align}

Putting it all together by \cref{thm:second_order_simulation}, we can simulate the spectrum of a target $XZ$-Hamiltonian $H_{target} = \sum_{ij} \frac{3}{8}\mu_{ij} (X_iX_j + Z_iZ_j)$ where $\mu_{ij} \geq 0\, \forall i\neq j$ using $k^{\text{th}}$-Laplacian of an independence complex where $k = n$ is the number of qubits in the target $XZ$-Hamiltonian. Since estimating the ground energy of $XZ$-Hamiltonian is QMA-hard, this completes the proof that estimating the minimum eigenvalue of the $k^{\text{th}}$-Laplacian of an independence complex is QMA-hard.

\section*{Acknowledgements}
C.R acknowledges that this material is based upon work supported by the U.S. Department of Energy, Office of Science, Accelerated Research in Quantum Computing, Fundamental Algorithmic Research toward Quantum Utility (FAR-Qu). C.R. thanks Ojas Parekh, Kevin Thompson, Jun Takahashi and Andrew Zhao for insightful discussions during the course of this work. C.R. also thanks Ojas Parekh for providing feedback on presentation and writing of this paper. 

\appendix
\section{Perturbative Gadgets}
\label{sec:perturbative_gadgets}
In this section we recall some perturbation theory based on Schrieffer–Wolff transformation \cite{bravyi2011schrieffer} required for us in the context of Hamiltonian simulation. Here the word \emph{simulation} is not to be confused with the simulation of Hamiltonian dynamics on a quantum computer. In our context, it means simulating the spectrum and eigenstates of a given Hamiltonian $H_{target}$ using another Hamiltonian $H_{sim}$. Formally, we use the notion of simulation as defined in \cite{bravyi2017complexity} which was later refined in \cite{cubitt2018universal}.

\begin{definition}[\cite{bravyi2017complexity}]\label{def:hamiltonian_simulations}
    Let $H_{target}$ be a Hamiltonian acting on a Hilbert space $\mathcal{H}_{\text{target}}$ of dimension $N$. A Hamiltonian $H_{sim}$ with a well-separated $N$-dimensional low-energy subspace $\mathcal{L}_N(H_{sim})$ and an isometry (encoding) $\mathcal{E}: \mathcal{H}_{target} \rightarrow \mathcal{H}_{sim}$ are said to simulate $H_{target}$ with an error $(\eta, \epsilon)$ if there exists an isometry $\tilde{\mathcal{E}}: \mathcal{H}_{target} \rightarrow \mathcal{H}_{\text{sim}}$ such that
    \begin{enumerate}
        \item The image of $\tilde{\mathcal{E}}$ coincides with the low-energy subspace $\mathcal{L}_N(H_{sim})$.
        \item $||H_{target} - \tilde{\mathcal{E}}^{\dagger}H_{\text{sim}}\tilde{\mathcal{E}}|| \leq \epsilon$.
        \item $||\mathcal{E}-\tilde{\mathcal{E}}|| \leq \eta$.
    \end{enumerate}
\end{definition}
Constructively, we can use perturbative gadgets at different orders to construct $H_{sim}$ that can simulate, as in \cref{def:hamiltonian_simulations}, a given $H_{target}$. We mention two theorems below one for simulation at first-order perturbation and another for second-order perturbation. Note that one can consider simulations at higher-order perturbations but we do not need them in this work.

Let $H_0$ be a Hamiltonian on Hilbert space $\mathcal{H}_{sim}$ with an $N$-dimensional ground subspace. Let $\Pi_{-}$ be the projector onto the ground subspace of $H_0$ and $\Pi_{+}$ be the projector onto the rest of the Hilbert space. For any given operator $O$ acting on $\mathcal{H}_{sim}$, let
\begin{align}
    O_{--} := \Pi_{-}O\Pi_{-}, \quad O_{-+} := \Pi_{-}O\Pi_{+}, \quad O_{+-} := \Pi_{+}O\Pi_{-}, \quad O_{++} := \Pi_{+}O\Pi_{+}.
\end{align}
Assume that the lowest eigenvalue of $H_0$ is equal to $0$ which implies $(H_0)_{--} =0$ and that the second lowest eigenvalue of $H_0$ is greater than or equal to 1.

\begin{theorem}[First-order simulation \cite{bravyi2017complexity}]
    Suppose one can choose an $H_0$, a $V$ and an isometry $\mathcal{E}$ such that
    \begin{align}
        ||\mathcal{E}H_{\text{target}} \mathcal{E}^{\dagger} - (V)_{--}|| \leq \epsilon/2.
    \end{align}
    Then $H_\text{sim} = \Delta H_0 + V$ simulates $H_{\text{target}}$ with an error $(\eta, \epsilon)$ provided that $\Delta \geq O(\epsilon^{-1}||V||^2 + \eta^{-1}||V||)$.
\end{theorem}

\begin{theorem}[Second-order simulation \cite{bravyi2017complexity}]
    \label{thm:second_order_simulation}
    Suppose one can choose $H_0$, $V_{main}$, $V_{extra}$ and an isometry $\mathcal{E}$ such that $V_{extra}$ is block diagonal with respect to $\Pi_{-}$ and $\Pi_{+}$, $(V_{main})_{--} = 0$ and
    \begin{align}
        ||\mathcal{E}H_{target} \mathcal{E}^{\dagger} - (V_{\text{extra}})_{--} + (V_{main})_{-+} H_0^{-1}(V_{main})_{+-}|| \leq \epsilon/2.
    \end{align}
    Suppose the norms of $V_{main}$, $V_{extra}$ is at most $\Lambda$. Then $H_{sim} = \Delta H_0 + \Delta^{1/2}V_{main} + V_{extra}$ simulates $H_{target}$ with an error $(\eta, \epsilon)$ provided that $\Delta \geq O(\epsilon^{-2}\Lambda^6 + \eta^{-2}\Lambda^2)$.
\end{theorem}
\noindent The condition in \cref{thm:second_order_simulation} that $V_{extra}$ should be block diagonal i.e., $(V_{extra})_{+-} = 0$  can be relaxed to $(V_{main})_{-+} H_0^{-1}(V_{extra})_{+-} = 0$ while the rest of the theorem still holds.

Note that the \cref{def:hamiltonian_simulations} of simulation does not restrict the Hilbert space $\mathcal{H}_{sim}$ to have underlying fermionic or a qubit structure. In the case of Fermionic Independent Set, we begin from a Hilbert space arising from fermionic modes but the independent set constraints define the accessible Hilbert space $\mathcal{H}_{sim}$.

\bibliographystyle{unsrt}
\bibliography{refs}

\end{document}